\documentclass[12pt]{article}
\def\lddots{\mathinner{\mkern1mu\raise1pt\hbox{.}\mkern2mu
    \raise4pt\hbox{.}\mkern2mu\raise7pt\vbox{\kern7pt\hbox{.}}\mkern1mu}}
\makeatletter
\def\numberbysection{\@addtoreset{equation}{section}
 	\def\theequation{\thesection.\arabic{equation}}}
\makeatother

\numberbysection

\newcommand{\be}{\begin{eqnarray}}
\newcommand{\ee}{\end{eqnarray}}
\newcommand{\non}{\nonumber}
\newcommand{\n}{\ensuremath{\mathcal{N}}}
\newcommand{\tr}{\mathop{\rm tr}\nolimits}

\begin{document}

\begin{titlepage}
\strut\hfill UMTG--207
\vspace{.5in}
\begin{center}

\LARGE Duality and quantum-algebra symmetry of the $A_{\n-1}^{(1)}$ open spin 
chain with diagonal boundary fields \\[1.0in]
\large Anastasia Doikou and Rafael I. Nepomechie\\[0.8in]
\large Physics Department, P.O. Box 248046, University of Miami\\[0.2in]  
\large Coral Gables, FL 33124 USA\\

\end{center}

\vspace{.5in}

\begin{abstract}
We show that the transfer matrix of the $A_{\n-1}^{(1)}$ open spin 
chain with diagonal boundary fields has the symmetry 
$U_{q}\left( SU(l) \right) \times U_{q}\left( SU(\n-l) \right) \times U(1)$, 
as well as a ``duality'' symmetry which maps $l \leftrightarrow \n - l$.
We exploit these symmetries to compute exact boundary $S$ matrices in the 
regime with $q$ real.
\end{abstract}

\end{titlepage}

\section{Introduction}

The richness of boundary phenomena is a dominant theme of contemporary 
theoretical physics.  Well-known examples include catalysis of baryon 
decay, D-branes, particle production near event horizons, the Kondo 
problem, and edge excitations in the fractional quantum Hall effect.  
As much of the boundary phenomena of interest is nonperturbative, 
integrable models with boundaries provide a particularly valuable 
laboratory for its study \cite{binder}, \cite{GZ}.  Magnetic chains 
associated with affine Lie algebras \cite{bazhanov}, \cite{jimbo} 
constitute large classes of such integrable models.  We focus here on 
the simplest such class, namely, the $A_{\n-1}^{(1)}$ spin chain.  In 
addition to its value as a simple toy model, this model is of interest 
in its own right: it is related to loop models describing certain 
self-avoiding walks (see, e.g., \cite{loop} and references therein), 
and possibly also to the $A_{\n-1}^{(1)}$ Toda field theory with 
imaginary coupling \cite{toda}, \cite{BL}.

Pasquier and Saleur observed \cite{PS} that a certain integrable open 
XXZ spin chain Hamiltonian has the quantum-algebra (or ``quantum 
group'') \cite{qgroups} symmetry $U_{q}\left( SU(2) \right)$.  Kulish 
and Sklyanin later showed \cite{KS} that this symmetry extends to the 
full transfer matrix \cite{sklyanin}.  This result was then 
generalized \cite{mn/nonsymmetric} - \cite{mn/addendum} to higher 
rank: namely, the transfer matrix of the $A_{\n-1}^{(1)}$ open spin 
chain {\it without} boundary fields has the symmetry $U_{q}\left( 
SU(\n) \right)$.  This model was further investigated in Refs.  
\cite{DVGR1},\cite{DVGR2}.  Moreover, the most general diagonal 
boundary interactions which preserve integrability were found 
\cite{DVGR0},\cite{DVGR3}.  However, the question of what symmetry -- 
if any -- remains in the presence of such boundary interactions was 
not explored.

We show here that by turning on diagonal boundary fields, the 
$U_{q}\left( SU(\n) \right)$ symmetry is broken to $U_{q}\left( SU(l) 
\right)$ $\times$ $U_{q}\left( SU(\n-l) \right)$ $\times$ $U(1)$.  Moreover, 
we find a ``duality'' symmetry which relates $l \leftrightarrow \n - 
l$.  We exploit these symmetries to compute exact boundary $S$ 
matrices \cite{GZ}, which describe the scattering of the model's 
excitations from the ends of the chain, in the regime with $q$ real.  
For the case $\n=2$, we recover the results of Refs.  \cite{JKKKM} and 
\cite{DMN2}; and for $q=1$ we recover the recent results \cite{DN}.

In Section 2 we review the construction of the open chain transfer 
matrix, and we exhibit its symmetries.  In Section 3 we summarize the 
computation of boundary $S$ matrices.  We conclude in Section 4 with a 
brief discussion of some possible generalizations of this work.  Some 
technical details have been relegated to the Appendices.

\section{The transfer matrix and its symmetries}

There are two basic building blocks for constructing integrable open 
spin chains:

\begin{enumerate}
	
\item The $R$ matrix, which is a solution of the Yang-Baxter equation 

\be
R_{12}(\lambda)\ R_{13}(\lambda + \lambda')\ R_{23}(\lambda')
= R_{23}(\lambda')\ R_{13}(\lambda + \lambda')\ R_{12}(\lambda) \,,
\label{YB}
\ee 
(see, e.g., \cite{QISM}).  We assume that the $R$ matrix has the 
unitarity property
\be
R_{12}(\lambda) R_{21}(-\lambda) = 1 \,, 
\label{property1}
\ee
where $R_{21}(\lambda) = {\cal P}_{12} R_{12}(\lambda) {\cal P}_{12} = 
R_{12}(\lambda)^{t_{1} t_{2}}$, $t$ denotes transpose, and ${\cal 
P}_{12}$ is the permutation matrix\footnote{The permutation matrix is 
defined by ${\cal P} x \otimes y = y \otimes x$ for all vectors $x$ 
and $y$ in an $\n$-dimensional complex vector space $C_{\n}.$} , and 
also the property (see also \cite{RSTS} )
\be
R_{12}(\lambda)^{t_{1}} M_{1} R_{12}(-\lambda - 2\rho)^{t_{2}} 
M_{1}^{-1} \propto 1 \,, 
\label{property2}
\ee
with $M^{t} = M$ and 
\be
\left[ M_{1} M_{2} \,, R_{12}(\lambda) \right] = 0 \,.
\label{property3}
\ee 

\item The matrices $K^{\pm}$, which are solutions 
of the boundary Yang-Baxter equation \cite{cherednik}
\footnote{We use here a $K^{+}$ which is related to the $K^{+}_{old}$ 
used in earlier work \cite{mn/nonsymmetric} - \cite{DVGR3} by
\be
K^{+}_{old}(\lambda) = M  K^{+}(-\lambda - \rho)^{t} \,.
\ee
The matrix $K^{+}_{old}$ satisfies an equation more complicated than 
(\ref{boundaryYB}), but the expression for the transfer matrix in 
terms of $K^{+}_{old}$ is simple.  The matrix $K^{+}$ which we use 
here obeys the same boundary Yang-Baxter equation as $K^{-}$, but the 
expression for the transfer matrix in terms of this $K^{+}$ (see Eq.  
(\ref{transfer1})) is more complicated.  The two approaches are of 
course equivalent.}
\be
R_{12}(\lambda_{1}-\lambda_{2})\ K_{1}^{\pm}(\lambda_{1})\
R_{21}(\lambda_{1}+\lambda_{2})\ K_{2}^{\pm}(\lambda_{2}) \non \\
= K_{2}^{\pm}(\lambda_{2})\ R_{12}(\lambda_{1}+\lambda_{2})\ 
K_{1}^{\pm}(\lambda_{1})\ R_{21}(\lambda_{1}-\lambda_{2}) \,.
\label{boundaryYB}
\ee 

\end{enumerate}

The corresponding transfer matrix $t(\lambda)$ for an open chain of 
$N$ spins is given by 
\cite{sklyanin}, \cite{mn/nonsymmetric}
\be
t(\lambda) = \tr_{0} M_{0}\ K_{0}^{+}(-\lambda-\rho)^{t}\  
T_{0}(\lambda)\  K^{-}_{0}(\lambda)\ \hat T_{0}(\lambda)\,,
\label{transfer1}
\ee
where $\tr_{0}$ denotes trace over the ``auxiliary space'' 0,
$T_{0}(\lambda)$ is the monodromy matrix 
\be
T_{0}(\lambda) = R_{0N}(\lambda) \cdots  R_{01}(\lambda) \,, 
\label{monodromy}
\ee
and 
$\hat T_{0}(\lambda)$ is given by
\be
\hat T_{0}(\lambda) = R_{10}(\lambda) \cdots  R_{N0}(\lambda) \,. 
\label{hatmonodromy}
\ee
(As is customary, we usually suppress the ``quantum-space'' subscripts 
$1 \,, \ldots \,, N$.) Indeed, it can be shown that this transfer 
matrix has the commutativity property
\be
\left[ t(\lambda)\,, t(\lambda') \right] = 0 \,.
\label{commutativity}
\ee 

In this paper, we consider the case of the $A_{\n-1}^{(1)}$ $R$ matrix 
\cite{devega}
\be
R_{12}(\lambda)_{j j \,, j j} &=& 1 \,, \non \\
R_{12}(\lambda)_{j k \,, j k} &=& {\sinh (-i \eta \lambda) 
\over \sinh \left( \eta (-i \lambda + 1) \right)} \,, \qquad j \ne k 
\,, \non \\
R_{12}(\lambda)_{j k \,, k j} &=& {\sinh \eta 
\over \sinh \left( \eta (-i \lambda + 1) \right)}e^{i \eta \lambda 
sign(j-k)} \,, \qquad j \ne k \,, \non \\
& & 1 \le j \,, k \le \n \,,
\label{Rmatrix} 
\ee
which depends on the anisotropy parameter $\eta \ge 0$, and which becomes
$SU(\n)$ invariant for $\eta \rightarrow 0$. This $R$ matrix has the properties
(\ref{property1}) and (\ref{property2}), with \cite{DVGR0}
\be
M_{j k} = \delta_{j k} e^{\eta (\n - 2 j + 1) }\,, \qquad \
\rho = i\n/ 2 \,.
\ee
Moreover, we consider the $\n \times \n$ diagonal $K$ matrices given by
\cite{DVGR3}
\be
K^{-}(\lambda) &=& K_{(l)}(\lambda \,, \xi_{-}) \,, \non \\ 
K^{+}(\lambda) &=& K_{(l)}(\lambda \,, \xi_{+} - {\n\over 2}) \,, 
\ee
where 
\be
K_{(l)}(\lambda \,, \xi) = diag \Bigl(  
\underbrace{a\,, \ldots \,, a}_{l} \,, 
\underbrace{b\,, \ldots \,, b}_{\n-l} \Bigr) \,, \non \\ 
a = \sinh \left( \eta( \xi + i \lambda ) \right) e^{- i \eta \lambda} \,, 
\qquad b = \sinh \left( \eta( \xi - i \lambda ) \right)  e^{ i \eta \lambda} \,,
\label{Kmatrix}
\ee
for arbitrary $\xi$, and any $l \in \{ 1 \,, \ldots \,, \n-1 \}$. 
Eq. (\ref{Kmatrix}) is the most 
general diagonal solution of the boundary Yang-Baxter equation
with the $A_{\n -1}^{(1)}$ $R$ matrix, up to an irrelevant overall factor. 

We shall usually denote by $t_{(l)}(\lambda \,, \xi_{-} \,, \xi_{+})$
the corresponding open spin chain transfer matrix 
\footnote{The more general transfer matrix 
$t_{(l_{+}, l_{-})}(\lambda\,, \xi_{-} \,, \xi_{+})$ constructed with 
$K_{(l_{\mp})}(\lambda \,, \xi)$ also forms a one-parameter 
commutative family.  For simplicity, we consider here (as in \cite{DN})
the special case $l_{+} = l_{-} = l$.}
\be
t_{(l)}(\lambda \,, \xi_{-} \,, \xi_{+}) = 
\tr_{0} M_{0}\ K_{(l)\ 0}(-\lambda-\rho \,, \xi_{+} - {\n\over 2})\  
T_{0}(\lambda)\  K_{(l)\ 0}(\lambda \,, \xi_{-})\ \hat T_{0}(\lambda)\,.
\label{transfer2}
\ee
The corresponding open spin chain Hamiltonian ${\cal H}_{open}$ is 
related to the derivative of the transfer matrix at $\lambda = 0$:
\be
{\cal H}_{open} &=& {i \over 4 \sinh (\eta \xi_{-}) 
\tr M K_{(l)}(-\rho \,, \xi_{+} - {\n\over 2})}  
{d\over d \lambda} t_{(l)}(\lambda\,, \xi_{-}\,, \xi_{+})
\Big\vert_{\lambda=0} \non \\ 
&\quad & - {i\over 4} 
{d\over d \lambda} \log \tr M K_{(l)}(-\lambda -\rho 
\,, \xi_{+} - {\n\over 2}) \Big\vert_{\lambda=0} \non  \\ 
&=& \sum_{n=1}^{N-1} {\cal H}_{n n+1} 
+ {i\over 4 \sinh (\eta \xi_{-})} {d\over d \lambda} 
K_{(l)\ 1}(\lambda \,, \xi_{-}) \Big\vert_{\lambda=0} \non  \\ 
&\quad & + {\tr_{0} M_{0} K_{(l)\ 0}(-\rho  \,, \xi_{+} - {\n\over 2})
{\cal H}_{N 0}\over \tr M K_{(l)}(-\rho \,, \xi_{+} - {\n\over 2})} 
\,,
\ee
where the two-site Hamiltonian ${\cal H}_{j k}$ is given by
\be
{\cal H}_{j k} = {i \over 2} {\cal P}_{j k} {d\over d \lambda} 
R_{j k}(\lambda) \Big\vert_{\lambda=0}  \,.
\ee
One can verify that the Hamiltonian is Hermitian.

The parameters $\xi_{\mp}^{-1}$ may be regarded as certain boundary 
fields.  For $\xi \rightarrow \infty$, the $K$ matrix (\ref{Kmatrix}) 
evidently becomes proportional to the identity matrix 
\be
K_{(l)}(\lambda \,, \xi) \sim 
{1\over 2}e^{\eta \xi}\ 1 \,, 
\ee
and so the transfer matrix is $U_{q}\left( 
SU(\n) \right)$ invariant \cite{mn/mpla}, \cite{mn/addendum}.
We now exhibit exact symmetries of this transfer matrix for finite 
values of $\xi_{\mp}$.

\subsection{Duality}

The duality symmetry of the transfer matrix is a remnant of the 
cyclic ($Z_{\n}$) symmetry \cite{devega}, \cite{belavin} of the 
$A_{\n-1}^{(1)}$ $R$ matrix. Indeed, the $R$ matrix (\ref{Rmatrix}) 
satisfies
\be
U_{1}\ R_{12}(\lambda)\ U_{1}^{-1} &=& V_{2}(-\lambda)^{-1}\
 R_{12}(\lambda)\ V_{2}(-\lambda) \non \\
U_{2}\ R_{12}(\lambda)\ U_{2}^{-1} &=& V_{1}(\lambda)^{-1}\
 R_{12}(\lambda)\ V_{1}(\lambda) \,,
\ee
where $U$ is the $\n \times \n$ matrix 
\be
U_{j k}= \delta_{j \,, k-1} + \delta_{\n \,, j} \delta_{1 \,, k}
\ee
which has the property $U^{\n} = 1$, and $V(\lambda)$ is the matrix
\be
V(\lambda)_{j k}= e^{-i \eta \lambda} \delta_{j \,, k-1} 
+ e^{i \eta \lambda} \delta_{\n \,, j} \delta_{1 \,, k} \,.
\ee
The corresponding quantum-space operator ${\cal U}$ defined by
\be
{\cal U} = U_{1}\ U_{2} \cdots U_{N} 
\label{calU}
\ee
therefore has the following action on the monodromy matrices
\be
{\cal U}\ T_{0}(\lambda)\ {\cal U}^{-1} 
&=&  V_{0}(\lambda)^{-1}\ T_{0}(\lambda)\ V_{0}(\lambda) \,, \non \\
{\cal U}\ \hat T_{0}(\lambda)\ {\cal U}^{-1} 
&=&  V_{0}(-\lambda)^{-1}\ \hat T_{0}(\lambda)\ V_{0}(-\lambda) \,,
\ee
and the transfer matrix (\ref{transfer2}) transforms as follows:
\be
{\cal U}^{l}\ t_{(l)}(\lambda\,, \xi_{-}\,, \xi_{+})\ {\cal U}^{-l} 
&=&  \tr_{0} \Big\{ \left( V(-\lambda)^{l}\ M 
K_{(l)}(-\lambda -\rho\,, \xi_{+} - {\n\over 2})\ V(\lambda)^{-l}
\right){}_{0} \non \\
&\times & T_{0}(\lambda) \left( V(\lambda)^{l}\ K_{(l)}(\lambda \,, \xi_{-})\ 
V(-\lambda)^{-l} \right){}_{0}\ \hat T_{0}(\lambda) \Big\} \,.
\ee

We now observe that the $K$ matrix (\ref{Kmatrix}) satisfies 
\be 
V(\lambda)^{l}\ K_{(l)}(\lambda \,, \xi_{-})\ V(-\lambda)^{-l} &=&
- e^{- 2 i (l - 1) \eta \lambda} K_{(\n - l)}(\lambda \,, -\xi_{-}) 
\,, \non \\ 
V(-\lambda)^{l}\ M K_{(l)}(-\lambda - \rho \,, \xi_{+} - {\n\over 2})\ 
V(\lambda)^{-l} &=&
- e^{2 i (l - 1) \eta \lambda} e^{\eta (\n - 2 l)} 
M K_{(\n - l)}(-\lambda - \rho \,, -\xi_{+} + {\n\over 2}) \,, \non \\ 
& &  l = 1\,, \ldots  \,, \n - 1 \,.
\ee

We conclude that the transfer matrix has the ``duality'' transformation 
property \footnote{The prefactor $e^{\eta (\n - 2l)}$ can be 
absorbed in the definition of $K^{+}$, and is not significant.}
\be
{\cal U}^{l}\ t_{(l)}(\lambda\,, \xi_{-} \,, \xi_{+})\ {\cal U}^{-l} 
= e^{\eta (\n - 2l)}\ t_{(l')}(\lambda\,, \xi'_{-} \,, \xi'_{+}) \,,
\label{duality}
\ee
where
\be
\xi'_{-} &=& -\xi_{-}  \,, \non \\
\xi'_{+} &=& -\xi_{+} + \n \,, \non \\ 
l' &=& \n -l \,.
\label{primes}
\ee
Notice that ${\cal U}^{l'} {\cal U}^{l} = 1$.  
The transfer matrix is ``self-dual'' for $\xi_{-} = \xi'_{-} = 0$,
$\xi_{+} = \xi'_{+} = \n/2$, and $l = l' = \n/2$.

\subsection{Quantum algebra symmetry}

In the defining representation of $SU(\n)$, we identify (following the 
notations of our previous paper \cite{DN}) the raising and lowering 
operators
\be 
j^{+ (k)} = e_{k, k+1} \,, \qquad
j^{- (k)} = e_{k+1, k} \,, \qquad k = 1 \,, \ldots \,, \n-1 \,,
\label{raising/lowering}
\ee
which correspond to the simple roots, and the Cartan generators
\be 
s^{(k)} = e_{k, k} - e_{k+1, k+1} \,, \qquad k = 1 \,, \ldots \,, \n-1 \,,
\label{cartan}
\ee 
where $e_{k,l}$ are elementary $\n \times \n$ matrices with matrix elements 
$\left( e_{k,l} \right)_{a b} = \delta_{k,a} \delta_{l,b}$.
We denote by $j^{\pm (k)}_{n}$, $s^{(k)}_{n}$ the generators at site $n$,
e.g.,
\be
s^{(k)}_{n} = 1 \otimes \ldots \otimes 1 \otimes 
\stackrel{\stackrel{n^{th}}{\downarrow}}{s^{(k)}} \otimes 1 
\otimes \ldots \otimes 1 \,, \qquad n = 1 \,, \ldots \,, N \,.
\ee
The corresponding generators of the quantum algebra $U_{q}\left( 
SU(\n) \right)$ which act on the full space of states are given by
(see, e.g., \cite{qgroups},\cite{DVGR2})
\be
J^{\pm (k)} &=& \sum_{n=1}^{N} q^{-s^{(k)}_{N}/2} \cdots 
q^{-s^{(k)}_{n+1}/2}\ j^{\pm (k)}_{n}\ q^{s^{(k)}_{n-1}/2}\ \cdots 
q^{s^{(k)}_{1}/2} \,,  
\non \\  
S^{(k)} &=& \sum_{n=1}^{N} s^{(k)}_{n} \,, \qquad k = 1 \,, \ldots \,, \n-1 \,.
\label{generators}
\ee 
These generators obey the commutation relations
\be
\left[ J^{+(k)} \,, J^{-(j)} \right] = \delta_{k,j} \left[ S^{(k)} 
\right]_{q} \,, 
\qquad 
\left[ S^{(k)} \,, J^{+(j)} \right] = \left( 2 \delta_{k,j} 
-  \delta_{k-1,j} - \delta_{k+1,j} \right) J^{+(j)} \,,
\label{algebra}
\ee 
where $\left[ x \right]_{q} \equiv ( q^{x} -  q^{-x})/(q - q^{-1})$.

We claim that the transfer matrix $t_{(l)}(\lambda \,, \xi_{-} \,, 
\xi_{+})$ has the invariance $U_{q}\left( SU(l) \right) \times 
U_{q}\left( SU(\n-l) \right) \times U(1)$.  In particular, we shall 
now show that
\be
\left[ t_{(l)}(\lambda \,, \xi_{-} \,, \xi_{+}) \,, S^{(k)} 
\right] &=& 
0 \,, \qquad k = 1 \,, \ldots \,, \n-1 \,, \label{first}  \\
\left[ t_{(l)}(\lambda \,, \xi_{-} \,, \xi_{+}) \,, J^{\pm (k)} 
\right] &=& 
0 \,, \qquad k \ne l \,, 
\label{second}
\ee 
where $q=e^{-\eta}$. Readers who are interested primarily in the 
calculation of boundary $S$ matrices may now wish to skip directly to 
Section 3, which may be read independently of the proof of Eqs. 
(\ref{first}) and (\ref{second}) that is presented below.

It is convenient to introduce, following Sklyanin \cite{sklyanin}, 
the quantity ${\cal T}_{(l)\ 0}(\lambda \,, \xi )$ defined by
\be
{\cal T}_{(l)\ 0}(\lambda \,, \xi ) = 
T_{0}(\lambda)\ K_{(l)\ 0}(\lambda \,, \xi )\ \hat T_{0}(\lambda) \,,
\label{calT}
\ee
in terms of which the expression (\ref{transfer2}) for the transfer 
matrix acquires the more compact form
\be
t_{(l)}(\lambda \,, \xi_{-} \,, \xi_{+}) = 
\tr_{0} M_{0}\ K_{(l)\ 0}(-\lambda-\rho \,, \xi_{+} - {\n\over 2})\ 
{\cal T}_{(l)\ 0}(\lambda \,, \xi_{-} )
\,.
\label{transfer3}
\ee

The proof of the first relation (\ref{first}) is straightforward. 
From the identity
\be
\left[ s_{2}^{(k)} \,, R_{12}(\lambda) \right] = 
- \left[ s_{1}^{(k)} \,, R_{12}(\lambda) \right] \,,
\ee
it follows that
\be
\left[ S^{(k)} \,, T_{0}(\lambda) \right] =
- \left[ s_{0}^{(k)} \,, T_{0}(\lambda) \right] \,,
\ee
and similarly for $\hat T_{0}(\lambda)$. Hence, 
\be
\left[ S^{(k)} \,, {\cal T}_{(l)\ 0}(\lambda \,, \xi ) \right] = 
- \left[ s_{0}^{(k)} \,, {\cal T}_{(l)\ 0}(\lambda \,, \xi ) \right] \,,
\ee
where we have also used the fact that 
$\left[ s_{0}^{(k)} \,, K_{(l)\ 0}(\lambda \,, \xi ) \right] = 0$. We 
conclude that
\be
\left[ S^{(k)} \,, t_{(l)}(\lambda \,, \xi_{-} \,, \xi_{+}) 
\right] &=& \tr_{0} \left[ S^{(k)} \,, M_{0}\ K_{(l)\ 0}
(-\lambda-\rho \,, \xi_{+} - {\n\over 2})\ 
{\cal T}_{(l)\ 0}(\lambda \,, \xi_{-} ) \right] \non \\ 
&=& - \tr_{0} \left[ s_{0}^{(k)} \,, M_{0}\ K_{(l)\ 0}
(-\lambda-\rho \,, \xi_{+} - {\n\over 2})\ 
{\cal T}_{(l)\ 0}(\lambda \,, \xi_{-} ) \right] \non \\
&=& 0 \,,
\ee
where the final equality follows from the cyclic property of the trace.

Our proof of the second relation (\ref{second}) relies on the well-known 
fact that in the limits $\lambda \rightarrow \pm i \infty$, the 
monodromy matrix becomes an upper/lower triangular matrix whose 
elements can be expressed in terms of the $U_{q}\left( SU(\n) 
\right)$ generators. (See, e.g., \cite{qgroups},\cite{mn/mpla},
\cite{DVGR2}.) It is therefore convenient to introduce  
notations for the corresponding limits of the $R$ and $T$ matrices:
\be
R_{12}(\lambda) & \rightarrow & e^{\mp \eta} R^{\pm}_{12} \,, \non \\
T_{0}(\lambda) & \rightarrow & e^{\mp \eta N} T^{\pm}_{0} \,, \non \\
\hat T_{0}(\lambda) & \rightarrow & e^{\mp \eta N} \hat T^{\pm}_{0} \,,
\ee
for $\lambda \rightarrow \pm i \infty$, with
\be
T^{\pm}_{0} &=& R^{\pm}_{0N} \cdots  R^{\pm}_{01} \,, \non \\
\hat T^{\pm}_{0} &=& R^{\pm}_{10}\cdots  R^{\pm}_{N0} \,.
\ee
Schematically, 
\be
T^{+}_{0} & \sim & \left(
           \begin{array}{ccccc}
           * & J^{-(1)} \\
		     & *                 &         & \mbox{ \LARGE *}     \\
			 &                   & \ddots                      \\
			 & \mbox{ \LARGE 0}  &         & *    & J^{-(\n - 1)} \\
	         &                   &         &      & * 
           \end{array} \right) \sim \hat T^{-}_{0} \,, \non \\ 
T^{-}_{0} & \sim & \left(
           \begin{array}{ccccc}
              *      \\
		   J^{+(1)}  & *                &          & \mbox{ \LARGE 0} \\
			         &                  &  \ddots          \\
			         & \mbox{ \LARGE *} &          & *    \\   
	                 &                  &          & J^{+(\n - 1)}   & * 
           \end{array} \right) \sim \hat T^{+}_{0} \,,
\label{schematic}
\ee
where the diagonal matrix elements involve only the Cartan generators.
We shall not need more explicit expressions for $T^{\pm}_{0}$ and 
$T^{\pm}_{0}$, which can be found e.g. in \cite{DVGR2}.

It is also important to realize that in these limits the $K$ 
matrices (\ref{Kmatrix}) become projectors:
\be
K_{(l)}(\lambda \,, \xi ) \sim  -{1\over 2}e^{\mp 2 i \eta 
\lambda - \eta \xi}\ \Pi_{(l)}^{\pm} 
\ee
for $\lambda \rightarrow \pm i \infty$, where $\Pi_{(l)}^{\pm}$ 
are the orthogonal projection operators
\be
\Pi_{(l)}^{+} &=& diag \Bigl(  
\underbrace{1\,, \ldots \,, 1}_{l} \,, 
\underbrace{0\,, \ldots \,, 0}_{\n-l} \Bigr) \,, \non \\
\Pi_{(l)}^{-} &=& diag \Bigl(  
\underbrace{0\,, \ldots \,, 0}_{l} \,, 
\underbrace{1\,, \ldots \,, 1}_{\n-l} \Bigr) \,,
\ee
and therefore $\Pi_{(l)}^{+} + \Pi_{(l)}^{-} = 1$.

It follows that for $\lambda \rightarrow \pm i \infty$, the quantity 
${\cal T}_{(l)\ 0}(\lambda \,, \xi )$ defined in Eq.  (\ref{calT}) 
tends (up to an irrelevant factor) to ${\cal T}_{(l)\ 0}^{\pm}$, where
\be
{\cal T}_{(l)\ 0}^{\pm} = T_{0}^{\pm}\ \Pi_{(l)\ 0}^{\pm}\ \hat 
T_{0}^{\pm}
\label{calTpm}
\,.
\ee 
It is easy to show (see Eq.  (\ref{OrthogSubspaces}) in Appendix A) 
that ${\cal T}_{(l)\ 0}^{+}$ and ${\cal T}_{(l)\ 0}^{-}$ obey
\be
\Pi_{(l)\ 0}^{\pm}\ {\cal T}_{(l)\ 0}^{\pm}\ \Pi_{(l)\ 0}^{\pm} =
{\cal T}_{(l)\ 0}^{\pm} \,.
\ee 
With the help of Eq.  (\ref{schematic}), one can also see that ${\cal 
T}_{(l)\ 0}^{+}$ depends on the generators $J^{\pm (k)}$ with $k < l$, 
while ${\cal T}_{(l)\ 0}^{-}$ depends on the generators $J^{\pm (k)}$ 
with $k > l$.

To prove the relation (\ref{second}), it suffices to show that the 
transfer matrix commutes with $\tau_{(l)}^{\pm}$
\be
\left[ t_{(l)}(\lambda \,, \xi_{-} \,, \xi_{+}) \,, \tau_{(l)}^{\pm}
\right] = 0 \,, 
\label{secondEquivalent}
\ee 
where $\tau_{(l)}^{\pm}$ are defined by
\be
\tau_{(l)}^{\pm} = \tr_{0} P_{0}^{\pm}\ {\cal T}_{(l)\ 0}^{\pm} \,,
\ee
and $P^{\pm}$ are $\n \times \n$ matrices which obey 
\be
\Pi_{(l)}^{\pm}\ P^{\pm}\ \Pi_{(l)}^{\pm} =
P^{\pm} \,,
\ee 
but which are otherwise arbitrary.  Indeed, with suitable choices of 
$P^{+}$ and $P^{-}$, one can project from ${\cal T}_{(l)\ 
0}^{+}$  and ${\cal T}_{(l)\ 0}^{-}$ the generators $J^{\pm (k)}$ with 
$k \ne l$. For example, with $P^{+} = j^{\pm (k)}$ one projects 
out $J^{\pm (k)}$ with $k < l$.

The quantities $\tau_{(l)}^{\pm}$ evidently have a structure similar 
to that of the transfer matrix.  The proof of Eq.  
(\ref{secondEquivalent}) is similar to the proof of the fundamental 
commutativity property (\ref{commutativity}) of the transfer matrix.  
Indeed, following Sklyanin, one can show using the properties 
(\ref{property2}), (\ref{property3}) that
\be
\tau_{(l)}^{\pm}\ t_{(l)}(\lambda \,, \xi_{-} \,, \xi_{+}) &=& 
\tr_{1} \left( P_{1}^{\pm} {\cal T}^{\pm}_{(l)\ 1} \right)^{t_{1}} \
\tr_{2} \left( M_{2} K_{(l)\ 2}(-\lambda-\rho \,, \xi_{+} - {\n\over 2}) 
{\cal T}_{(l)\ 2}(\lambda \,, \xi_{-} ) \right) \non \\
& \vdots & \non \\
&=& \tr_{12} \Big\{ R^{\mp}_{12} P_{1}^{\pm \ t_{1}} 
M_{1}^{-1} R_{21}^{\mp} K_{(l)\ 2}(-\lambda-\rho \,, \xi_{+} - {\n\over 
2})  \non \\ 
& & \times \left[ R^{\pm}_{12} {\cal T}^{\pm}_{(l)\ 1} R^{\pm}_{21} 
{\cal T}_{(l)\ 2}(\lambda \,, \xi_{-} ) M_{1} M_{2} \right]^{t_{12}} 
\Big\} \,.
\label{long1}
\ee
Similarly,
\be
t_{(l)}(\lambda \,, \xi_{-} \,, \xi_{+})\ \tau_{(l)}^{\pm} &=& 
\tr_{2} \left( M_{2} K_{(l)\ 2}(-\lambda-\rho \,, \xi_{+} - {\n\over 2}) 
{\cal T}_{(l)\ 2}(\lambda \,, \xi_{-} ) \right)\
\tr_{1} \left( P_{1}^{\pm} {\cal T}^{\pm}_{(l)\ 1} \right)^{t_{1}} \non \\
& \vdots & \non \\
&=& \tr_{12} \Big\{ K_{(l)\ 2}(-\lambda-\rho \,, \xi_{+} - 
{\n\over 2}) R^{\mp}_{12} P_{1}^{\pm \ t_{1}} M_{1}^{-1} R_{21}^{\mp}  
\non \\ 
& & \times \left[ {\cal T}_{(l)\ 2}(\lambda \,, \xi_{-} ) 
R^{\pm}_{12} {\cal T}^{\pm}_{(l)\ 1} R^{\pm}_{21} M_{1} M_{2} \right]^{t_{12}} 
\Big\}
\,. \label{long2}
\ee
It is easy to see that the quantities in square brackets in Eqs. 
(\ref{long1}) and (\ref{long2}) are equal. Indeed, since 
${\cal T}_{(l)}(\lambda \,, \xi )$ obeys the boundary 
Yang-Baxter equation
\be
R_{12}(\lambda_{1}-\lambda_{2})\ {\cal T}_{(l)\ 1}(\lambda_{1} \,, 
\xi )\
R_{21}(\lambda_{1}+\lambda_{2})\ {\cal T}_{(l)\ 2}(\lambda_{2} \,, 
\xi ) \non \\
= {\cal T}_{(l)\ 2}(\lambda_{2} \,, \xi )\ R_{12}(\lambda_{1}+\lambda_{2})\
{\cal T}_{(l)\ 1}(\lambda_{1} \,, \xi )\ R_{21}(\lambda_{1}-\lambda_{2}) \,,
\ee 
by taking $\lambda_{1} \rightarrow \pm i \infty$ we obtain the 
relations
\be
R_{12}^{\pm}\ {\cal T}_{(l)\ 1}^{\pm}\
R_{21}^{\pm}\ {\cal T}_{(l)\ 2}(\lambda \,, \xi ) =
{\cal T}_{(l)\ 2}(\lambda \,, \xi ) R_{12}^{\pm}\
{\cal T}_{(l)\ 1}^{\pm}\  R_{21}^{\pm}\ \,.
\ee 

In order to prove the commutativity 
(\ref{secondEquivalent}), it therefore suffices to show that
\be
\tr \left( A C \right) = \tr \left( B C \right) \,,
\ee
where
\be
A &=& R_{12}^{\mp} P_{1}^{\pm t_{1}} M_{1}^{-1} R_{21}^{\mp} 
K_{(l)\ 2}(-\lambda-\rho \,, \xi_{+} - {\n\over 2}) \,,
\non \\
B &=& K_{(l)\ 2}(-\lambda-\rho \,, \xi_{+} - {\n\over 2}) 
R^{\mp}_{12} P_{1}^{\pm \ t_{1}} M_{1}^{-1} R_{21}^{\mp}  \,, \non \\
C &=& \left[ R_{12}^{\pm}\ {\cal T}_{(l)\ 1}^{\pm}\
R_{21}^{\pm}\ {\cal T}_{(l)\ 2}(\lambda \,, \xi_{-} ) M_{1} M_{2}
\right]^{t_{12}} \,.
\ee
We establish this result by showing that $A-B$ and $C$ are trace 
orthogonal matrices. That is, we show that for some projection 
operator $\Pi$,
\be
\Pi\ C\ \Pi &=& C \non \\
\Pi\ \left( A - B \right)\ \Pi &=& 0 \,,
\ee
from which it immediately follows that
\be
\tr \left[ ( A - B )\ C \right] =\tr \left[ ( A - B )\ \Pi\ C\ \Pi \right] 
= \tr \left[ \Pi\ ( A - B )\ \Pi\ C \right] = 0 \,.
\ee
Indeed, in Appendix A we show that
\be
\Pi_{(l)\ 1}^{\pm} \left( R_{12}^{\pm}\ {\cal T}_{(l)\ 1}^{\pm}\
R_{21}^{\pm}\ {\cal T}_{(l)\ 2}(\lambda \,, \xi ) M_{1} M_{2} \right) 
\Pi_{(l)\ 1}^{\pm} &=&
R_{12}^{\pm}\ {\cal T}_{(l)\ 1}^{\pm}\
R_{21}^{\pm}\ {\cal T}_{(l)\ 2}(\lambda \,, \xi ) M_{1} M_{2} \,, 
\label{finalFirst} \\
\Pi_{(l)\ 1}^{\pm} \left( R_{12}^{\mp} P_{1}^{\pm t_{1}} M_{1}^{-1} 
R_{21}^{\mp} K_{(l)\ 2}(\lambda \,, \xi) \right) \Pi_{(l)\ 1}^{\pm} 
&=&
\Pi_{(l)\ 1}^{\pm} \left( K_{(l)\ 2}(\lambda \,, \xi) 
R^{\mp}_{12} P_{1}^{\pm \ t_{1}} M_{1}^{-1} R_{21}^{\mp} \right) 
\Pi_{(l)\ 1}^{\pm} \,, \non \\ 
\label{finalSecond}
\ee 
thereby completing the proof.

We have considered here symmetry transformations which leave the 
anisotropy parameter $\eta$ unchanged.  In Appendix C we briefly 
discuss related transformations, which however transform $\eta 
\rightarrow -\eta$.

\section{Boundary $S$ Matrix}

The eigenstates of the transfer matrix 
$t_{(l)}(\lambda \,, \xi_{-} \,, \xi_{+})$ can be constructed by the 
so-called nested Bethe Ansatz using the pseudovacuum $\omega_{(k)}$ 
defined by
\be
\omega_{(k)} = \underbrace{v_{(k)}\otimes \cdots \otimes v_{(k)}}_{N} \,,
\qquad k = 1\,, \ldots \,, \n \,, 
\ee
where $v_{(k)}$ are the standard $\n$-dimensional Cartesian basis 
vectors with components $v_{(k)\ a} = \delta_{k\,, a}$.  The 
corresponding Bethe Ansatz equations (BAE) can be cast in the 
form\footnote{We restrict to real $\lambda_\alpha^{(j)}$, which 
suffices for computing boundary $S$ matrices.}
\be
h^{(j)}_{(l)} ( \lambda_\alpha^{(j)} ) = J_\alpha^{(j)}  \,, 
\qquad \alpha = 1\,, \ldots \,, M^{(j)} \,, 
\qquad j = 1 \,, \ldots \,, \n-1 \,,
\label{BAlogopen} 
\ee 
where $h^{(j)}_{(l)}(\lambda)$ is the so-called counting function 
(explicit expressions are given below), and 
$\{ J_\alpha^{(j)} \}$ are integers lying in certain ranges which 
serve as ``quantum numbers'' of the Bethe Ansatz (BA) states. We label the 
holes in the ranges by $\{ \tilde J_\alpha^{(j)} \} \,, \quad 
\alpha = 1 \,, \ldots \,, \nu^{(j)}$. The corresponding hole rapidities
$\{ \tilde\lambda_{\alpha}^{(j)} \}$ are defined by
\be
h^{(j)}_{(l)}(\tilde\lambda_{\alpha}^{(j)}) =  \tilde J_{\alpha}^{(j)} 
\,, \qquad \alpha = 1 \,, \ldots \,, \nu^{(j)} \,.
\ee

The nature of the ground state and excitations is similar to that of 
the corresponding isotropic model \cite{faddeev/takhtajan}, 
\cite{su(n)},\cite{DN}, except that Lie algebra symmetries are now 
replaced by their $q$-deformations.  Indeed, the ground state is the 
BA state with no holes, i.e., with $\n-1$ filled Fermi seas.  
Moreover, the BA state with one hole in the $j^{th}$ sea (i.e., 
$\nu^{(k)}=\delta_{k \,, j}$ for $k = 1\,, \ldots \,, \n -1$) is a 
particle-like excited state which belongs \footnote{Since the boundary 
interactions break the bulk 
$U_{q}\left( SU(\n) \right)$ symmetry, states should -- strictly 
speaking -- be classified according to the unbroken symmetry 
$U_{q}\left( SU(l) \right) \times U_{q}\left( SU(\n-l) \right) \times 
U(1)$.  However, one expects that the effects of the boundary should 
be ``small'' at points of the chain that are far from the boundary.  
In particular, bulk multiparticle states should ``approximately'' form 
irreducible representations of $U_{q}\left( SU(\n) \right)$, and 
therefore we refer to such states by $U_{q}\left( SU(\n) \right)$ 
quantum numbers.} to the fundamental representation $[j]$ of 
$U_{q}\left( SU(\n) \right)$, corresponding to a Young tableau with a 
single column of $j$ boxes, as shown in Figure \ref{fig1}.
\setlength{\unitlength}{12pt}
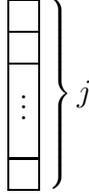
\begin{figure}[htb]
\centering
\be  
\left.
\begin{picture}(1,3.4)(0,2.8)
\put(0,0){\line(0,1){6}}
\put(1,0){\line(0,1){6}}
\put(0,0){\line(1,0){1}}
\put(0,1){\line(1,0){1}}
\put(0,4){\line(1,0){1}}
\put(0,5){\line(1,0){1}}
\put(0,6){\line(1,0){1}}
\put(.5,3.5){\makebox(0,0)[t]{$\vdots$}}
\end{picture}
\; \right\} j \non 
\ee 
\caption[xxx]{\parbox[t]{.7\textwidth}{%
 	   Young tableau with a	single column of $j$ boxes,	
 	   corresponding to	the	fundamental	representation 
	   $[j]$ of	$U_{q}\left( SU(\n) \right)$}
 	   }
 	   \label{fig1}
\end{figure}
This excitation carries energy $s^{(j)}(\tilde\lambda^{(j)})$ 
and ``momentum'' $p^{(j)}(\tilde\lambda^{(j)})$, where 
\be
{d\over d\lambda}p^{(j)}(\lambda) = 2 \pi s^{(j)}(\lambda) \,,
\ee
and the Fourier transform of $s^{(j)}(\lambda)$ is given 
by\footnote{Our conventions for Fourier transforms are as follows:
\be
f(\lambda) = {\eta \over \pi} \sum_{k=-\infty}^\infty  e^{-2 i \eta k \lambda}
\hat f(k) \,, \qquad
\hat f(k) = \int_{-\pi/2\eta}^{\pi/2\eta}d\lambda\ e^{2 i \eta k \lambda}
f(\lambda) \,.
\label{fourier1}
\ee}
\be
\hat s^{(j)}(k) = {\sinh \left( (\n - j) \eta |k| \right)\over
\sinh \left(\n \eta |k| \right)} \,.
\label{s}
\ee
Introducing the lattice spacing $a$, and taking the continuum 
($a \rightarrow 0$) and isotropic ($\eta \rightarrow 0$) limits with 
${4\over a}e^{-{\pi^{2}\over \n \eta}} = \mu $ finite, one obtains \cite{contlim} 
a relativistic dispersion relation, with mass $m^{(j)} = \mu \sin 
\left( \pi j/\n \right)$.

We define the boundary $S$ matrices $S^{\mp}_{(l)\ [j]}$ for a 
particle of type $[j]$ by the quantization condition \cite{GMN}
\be
\left( e^{i 2 p^{(j)}(\tilde\lambda^{(j)}) N} 
S^{+}_{(l)\ [j]}\ S^{-}_{(l)\ [j]}- 1 \right) 
|\tilde\lambda^{(j)} \rangle = 0 \,.
\label{quantizationopen}
\ee 
We compute these $S$ matrices from the ${1\over N}$ terms in the 
distribution of roots of the BAE for the corresponding BA states 
$|\tilde\lambda^{(j)} \rangle$.  Indeed, we rely on the key identity
\be 
{1\over \pi} {d\over d\lambda}p^{(j)}(\lambda) 
+ \sigma^{(j)}_{(l)}(\lambda) - 2 s^{(j)}(\lambda)  = 
{1\over N} {d\over d\lambda}h^{(j)}_{(l)}(\lambda) \,,
\label{key}
\ee
where the ``density'' $\sigma^{(j)}_{(l)}(\lambda)$ defined by
\be
\sigma^{(j)}_{(l)}(\lambda) = 
{1\over N} {d \over d\lambda} h^{(j)}_{(l)}(\lambda)
\label{definesigmaopen}
\ee
describes the distribution of roots of the BAE.  This method 
\cite{GMN} is a generalization of the approach pioneered by Korepin 
\cite{korepin} and Andrei-Destri \cite{andrei/destri} for calculating 
bulk two-particle $S$ matrices.  (See also \cite{FS}.)

In order to compute each of the boundary $S$ matrix elements, 
we shall need the densities $\sigma^{(j)}_{(l)}(\lambda)$ corresponding 
to each of the pseudovacua $\omega_{(1)} \,, \ldots \,, 
\omega_{(\n)}$. It can be shown that the Bethe Ansatz equations 
(and hence, the counting functions $h^{(j)}_{(l)}(\lambda)$ and densities
$\sigma^{(j)}_{(l)}(\lambda)$) corresponding to the first $l$ pseudovacua 
$\omega_{(1)} \,, \ldots \,, \omega_{(l)}$ are all the same; and 
the BAE corresponding to the last $\n - l$ pseudovacua 
$\omega_{(l+1)} \,, \ldots \,, \omega_{(\n)}$ are all the same.
This result is quite plausible, in view of the $U_{q}\left( SU(l) 
\right)$ $\times$ $U_{q}\left( SU(\n-l) \right)$ symmetry of the 
transfer matrix. For completeness, we provide in Appendix B a  
proof of this result.

Corresponding to the first $l$ pseudovacua 
$\omega_{(1)} \,, \ldots \,, \omega_{(l)}$, the BAE for the transfer 
matrix $t_{(l)}(\lambda \,, \xi_{-} \,, \xi_{+})$  are given by \cite{DVGR3}
\be
1 &=& \left[ e_{2\xi_{-} + l} (\lambda_{\alpha}^{(l)})\ 
e_{-\left( 2\xi_{+} - 2 \n + l \right)}(\lambda_{\alpha}^{(l)})\  
\delta_{l,j} + \left( 1 - \delta_{l,j} \right) \right]  \non \\  
& & \times \prod_{\beta=1}^{M^{(j-1)}} 
e_{-1}(\lambda_{\alpha}^{(j)} - \lambda_{\beta}^{(j-1)})\ 
e_{-1}(\lambda_{\alpha}^{(j)} + \lambda_{\beta}^{(j-1)})
\prod_{\stackrel{\scriptstyle\beta=1}{\scriptstyle\beta \ne \alpha}}^{M^{(j)}} 
e_{2}(\lambda_{\alpha}^{(j)} - \lambda_{\beta}^{(j)})\
e_{2}(\lambda_{\alpha}^{(j)} + \lambda_{\beta}^{(j)}) \non \\ 
& & \times \prod_{\beta=1}^{M^{(j+1)}} 
e_{-1}(\lambda_{\alpha}^{(j)} - \lambda_{\beta}^{(j+1)})\
e_{-1}(\lambda_{\alpha}^{(j)} + \lambda_{\beta}^{(j+1)}) \non \\ 
& &  \qquad \qquad 
\alpha = 1 \,, \ldots \,, M^{(j)} \,, \qquad  
j = 1\,, \ldots \,, \n-1  \,,
\label{openBAE}
\ee
where 
\be
e_n(\lambda) = 
{\sin  \eta \left( \lambda + {in\over 2} \right) 
\over           
 \sin  \eta \left( \lambda - {in\over 2} \right) } \,,
\ee
and $M^{(0)} = N \,, \quad M^{(\n)} = 0 \,, \quad 
\lambda_{\alpha}^{(0)} = \lambda_{\alpha}^{(\n)} = 0 \,.$ 
The requirement that solutions of the BAE 
correspond to independent BA states leads to the 
restriction $0 < \lambda_{\alpha}^{(j)} < {\pi\over 2 \eta}$.
For simplicity, we restrict to ``weak'' boundary fields $\xi_{-} > 
{1\over 2}(\n -1)$, $\xi_{+} > \n - {1\over 2}$. The 
energy as a function of $\xi_{-}$ and $\xi_{+}$ is discussed for the 
case $\n = 2$ in \cite{JKKKM}, \cite{KapSko}.

Defining the counting function so that the Bethe Ansatz equations take 
the form (\ref{BAlogopen}), we find
\be
h^{(j)}_{(l)}(\lambda) &=& {1\over 2\pi} \Bigl\{ q_{1}(\lambda) +
r_{1}(\lambda) +
\left[ -q_{2\xi_{-}+l}(\lambda)  + q_{2\xi_{+} - 2\n + l}(\lambda) 
\right] \delta_{j,l} \non \\ 
&+& \sum_{\beta=1}^{M^{(j-1)}} \left[ 
q_{1} (\lambda - \lambda_{\beta}^{(j-1)}) 
+ q_{1} (\lambda + \lambda_{\beta}^{(j-1)}) \right]
- \sum_{\beta=1}^{M^{(j)}} \left[ 
q_{2} (\lambda - \lambda_{\beta}^{(j)}) 
+ q_{2} (\lambda + \lambda_{\beta}^{(j)}) \right] \non \\ 
&+& \sum_{\beta=1}^{M^{(j+1)}} \left[ 
q_{1} (\lambda - \lambda_{\beta}^{(j+1)}) 
+ q_{1} (\lambda + \lambda_{\beta}^{(j+1)}) \right]
\Bigr\} \,, 
\label{countingopen} 
\ee 
where $q_n (\lambda)$ and $r_n (\lambda)$ are odd monotonic-increasing 
functions defined (for $n > 0$) by
\be
q_n (\lambda) &=& \pi + i\log e_n(\lambda) \,, \non \\ 
 -\pi < q_n (\lambda) & \le & \pi \quad \mbox{for} \quad 
-{\pi\over 2 \eta} < \lambda \le {\pi\over 2 \eta} \,, \quad \mbox{and} \quad  
q_{n}(\lambda + {\pi\over \eta}) = q_{n}(\lambda) + 2 \pi  \,, \\ 
r_n (\lambda) &=&  i\log g_n(\lambda) \,,
\qquad
g_n(\lambda) = 
{\cos  \eta \left( \lambda + {in\over 2} \right) 
\over           
 \cos  \eta \left( \lambda - {in\over 2} \right) } \,, \non \\ 
-\pi < r_n (\lambda) & \le & \pi \quad \mbox{for} \quad 
-{\pi\over 2 \eta} < \lambda \le {\pi\over 2 \eta} \,, \quad \mbox{and} \quad   
r_{n}(\lambda + {\pi\over \eta}) = r_{n}(\lambda) + 2 \pi  \,.
\ee

We consider now a multi-hole state, with the number of holes in each 
of the seas given by $\nu^{(1)} \,, \ldots \,, \nu^{(\n-1)}$.
In the thermodynamic limit, we obtain a system of linear integral 
equations for the densities (\ref{definesigmaopen}) by approximating 
the sums in $h^{(j)}_{(l)}(\lambda)$ by integrals using \cite{DMN2}
\be
{1\over N} \sum_{\alpha=1}^{M^{(j)}} g(\lambda_\alpha^{(j)}) \approx 
\int_{0}^{\pi/ 2\eta}  g(\lambda')\ \sigma^{(j)}_{(l)}(\lambda')\ d\lambda' 
- {1\over N} \sum_{\alpha=1}^{\nu^{(j)}} g(\tilde\lambda_\alpha^{(j)}) 
- {1\over 2N} \left[g(0) + g({\pi\over 2\eta}) \right]   \,.
\label{approximationopen} 
\ee
In this way we obtain
\be
& &\sum_{m=1}^{\n-1} \left( \left( \delta + {\cal K} \right)_{jm} * 
\sigma^{(m)}_{(l)\ s}\right) (\lambda) = 2 a_{1}(\lambda) \delta_{j,1} 
+ {1\over N} \Big\{ a_{2}(\lambda) + b_{2}(\lambda) \non \\
&+&  \left( a_{1}(\lambda) +  b_{1}(\lambda) \right) 
\left( -1 + \delta_{j,1} + \delta_{j,\n -1} \right) 
+\left( - a_{2\xi_{-} + l}(\lambda) 
+ a_{2\xi_{+} - 2 \n + l}(\lambda) \right) \delta_{j,l} \non \\
&+& \sum_{m=1}^{\n-1} \sum_{\alpha=1}^{\nu^{(m)}} 
\left( {\cal K}(\lambda - \tilde\lambda_\alpha^{(m)})_{jm} +
{\cal K}(\lambda + \tilde\lambda_\alpha^{(m)})_{jm} \right) \Big\} \,, 
\qquad j = 1\,, \ldots \,, \n -1 \,, 
\ee
where $\sigma^{(j)}_{(l)\ s}(\lambda)$ is the symmetric density defined by
\be
\sigma^{(j)}_{(l)\ s}(\lambda) = 
\left\{ \begin{array}{cc}
           \sigma^{(j)}_{(l)}(\lambda)   & \lambda > 0 \\
           \sigma^{(j)}_{(l)}(-\lambda)  & \lambda < 0   
         \end{array} \right. \,,
\ee   
$*$ denotes the convolution
\be
\left( f * g \right) (\lambda) = \int_{-\pi/ 2\eta}^{\pi/ 2\eta}
f(\lambda - \lambda')\ g(\lambda')\ d\lambda' \,, 
\ee  
and 
\be
{\cal K}(\lambda)_{jm} 
&=& a_{2}(\lambda) \delta_{m,j} 
- a_{1}(\lambda) (\delta_{m,j-1} + \delta_{m,j+1}) \,, \non \\
a_n(\lambda) &=& {1\over 2\pi} {d q_n (\lambda)\over d\lambda} 
= {\eta\over \pi} {\sinh (\eta n)\over \cosh(\eta n) - \cos (2 \eta 
\lambda)} \,, \non \\
b_n(\lambda) &=& {1\over 2\pi} {d r_n (\lambda)\over d\lambda} 
= {\eta\over \pi} {\sinh (\eta n)\over \cosh(\eta n) + \cos (2 \eta 
\lambda)} = a_{n}(\lambda \pm {\pi\over 2 \eta}) \,. 
\label{KK}
\ee
Using Fourier transforms, we obtain the solution
\be
\sigma^{(j)}_{(l)\ s}(\lambda) &=& 2 s^{(j)}(\lambda) 
+ \delta \sigma^{(j)}_{(l)}(\lambda) \non \\
&+& {1\over N}  \Big\{ \sum_{m=1}^{\n-1} 
\left( {\cal R}_{jm} * \left[ a_{2} + b_{2}
+ \left( a_{1} +  b_{1} \right)
\left( -1 + \delta_{m,1} + \delta_{m,\n -1} \right) \right] 
\right)(\lambda) \non  \\
&+& \sum_{m=1}^{\n-1} \sum_{\alpha=1}^{\nu^{(m)}}
\left[ \delta(\lambda - \tilde\lambda_{\alpha}^{(m)}) \delta_{j,m}
- {\cal R}_{jm}(\lambda - \tilde\lambda_{\alpha}^{(m)}) 
+( \tilde\lambda_{\alpha}^{(m)} \rightarrow 
- \tilde\lambda_{\alpha}^{(m)} ) \right] \Big\} \,,
\label{sigmaopen}
\ee 
where ${\cal R}_{jm}(\lambda)$ has the Fourier transform (see, e.g., 
\cite{devega})
\be
\hat {\cal R}_{m m'}(k) = {e^{\eta |k|} 
\sinh \left( m_{<} \eta |k| \right) 
\sinh \left( (\n - m_{>}) \eta |k|\right) \over
\sinh \left( \n \eta |k| \right) 
\sinh \left( \eta |k| \right)} 
\,,
\label{inverse}
\ee 
with $m_{>}=\max(m \,, m')$ and $m_{<}=\min(m \,, m')$, and 
\be 
\delta \sigma^{(j)}_{(l)}(\lambda) = {1\over N} 
\left( {\cal R}_{jl} * \left( - a_{2\xi_{-} + l}
+ a_{2\xi_{+} - 2 \n + l} \right) \right) (\lambda)
\label{deltasigma}
\ee
has the dependence on the boundary parameters $\xi_{\mp}$. We note also
\be
\hat a_{n}(k) = e^{-\eta n |k|} \,, \qquad 
\hat b_{n}(k) = (-)^{k}\ \hat a_{n}(k) \,, \qquad n > 0 \,. 
\ee 

The above density $\sigma^{(j)}_{(l)\ s}(\lambda)$ corresponds to the 
first $l$ pseudovacua $\omega_{(1)} \,, \ldots \,, \omega_{(l)}$.  We 
also need the density $\sigma'^{(j)}_{(l)\ s}(\lambda)$ corresponding 
to the last $\n - l$ pseudovacua $\omega_{(l+1)} \,, \ldots \,, 
\omega_{(\n)}$, as already remarked.  We recall that under duality, 
the transfer matrix transforms as (\ref{duality})
\be
{\cal U}^{l}\ t_{(l)}(\lambda\,, \xi_{-} \,, \xi_{+})\ {\cal U}^{-l} 
\propto t_{(l')}(\lambda\,, \xi'_{-} \,, \xi'_{+}) \,,
\ee
while the pseudovacuum $\omega_{(k)}$ transforms as
\be
{\cal U}^{l}\ \omega_{(k)} = \omega_{(k-l)} \,,
\label{dualvac}
\ee 
where we identify $\omega_{(k)} \equiv \omega_{(k + \n)} $. 
It follows \cite{DN} that the BAE for 
$t_{(l)}(\lambda\,, \xi_{-} \,, \xi_{+})$ with 
pseudovacua $\omega_{(l+1)} \,, \ldots \,, \omega_{(\n)}$ are 
the same as the BAE for 
$t_{(l')}(\lambda\,, \xi'_{-} \,, \xi'_{+})$ with 
$\omega_{(1)} \,, \ldots \,, \omega_{(l')}$; i.e., the BAE
are given by Eq.  (\ref{openBAE}), except 
with $\xi_{\mp} \rightarrow \xi'_{\mp}$ and $l \rightarrow l'$.  It 
follows that the corresponding densities $\sigma'^{(j)}_{(l)\ 
s}(\lambda)$ are given by Eq.  (\ref{sigmaopen}), except with
\be
\delta \sigma'^{(j)}_{(l)}(\lambda) = {1\over N} 
\left( {\cal R}_{j,\n -l} * \left(  a_{2\xi_{-} - \n + l}
- a_{2\xi_{+} - \n + l} \right) \right) (\lambda)
\label{deltasigmaprime}
\,. 
\ee

For simplicity, we compute boundary $S$ matrices only for the cases 
$[1]$ and $[\n-1] = [\bar 1]$, for which the $S$ matrices act in the 
$\n$-dimensional complex vector spaces $C_{\n}$ and 
$\overline{C_{\n}}$, respectively.

We begin with the case $[1]$, for which the boundary $S$ matrices 
$S^{\mp}_{(l)\ [1]}$ are diagonal $\n \times \n$ matrices of the form
\be
S^{\mp}_{(l)\ [1]} = diag \Bigl(  
\underbrace{\alpha^{\mp}_{(l)}\,, \ldots \,, \alpha^{\mp}_{(l)}}_{l} \,, 
\underbrace{\beta^{\mp}_{(l)}\,, \ldots \,, \beta^{\mp}_{(l)}}_{\n-l} \Bigr) \,.
\label{form[1]}
\ee
Choosing the state $|\tilde\lambda^{(1)} \rangle$ in Eq.  
(\ref{quantizationopen}) to be the BA state constructed with 
any of the first $l$ pseudovacua $\omega_{(1)} \,, \ldots \,, \omega_{(l)}$
having one hole in sea 1 (i.e.,
$\nu^{(j)} = \delta_{j\,, 1}$), it follows 
from the identity (\ref{key}) that (up to a rapidity-independent phase 
factor)
\be
\alpha^{+}_{(l)}\ \alpha^{-}_{(l)}  \sim   
\exp \left\{ i 2\pi N \int_{0}^{\tilde\lambda^{(1)}}
\left( \sigma_{(l)}^{(1)}(\lambda) - 2 s^{(1)}(\lambda) \right) d\lambda
\right\} \,,
\label{openrel}
\ee 
where $\sigma_{(l)}^{(1)}(\lambda)$ is given by Eq. (\ref{sigmaopen}) 
with $j=1$. We now observe that
\be
\int_{0}^{\tilde\lambda^{(1)}} \left[ 
{\cal R}\left(\lambda - \tilde\lambda^{(1)}\right) 
+ {\cal R}\left(\lambda + \tilde\lambda^{(1)}\right) \right]\ d\lambda 
= \int_{0}^{\tilde\lambda^{(1)}} 2 {\cal R}\left(2 \lambda \right)\ d\lambda 
\,.
\ee
Moreover, we note the identity 
\be
\sum_{k=1}^{\infty}{\left( 1 - e^{-2 \eta k \beta} \right) 
\left( 1 - e^{-2 \eta k \gamma} \right) e^{-2 \eta k \mu}\over 1 - 
e^{- 2 \eta k}}  
{1\over k} = \log {\Gamma_{q^{2}}(\mu)
\Gamma_{q^{2}}(\mu + \beta + \gamma)\over 
\Gamma_{q^{2}}(\mu + \beta)\Gamma_{q^{2}}(\mu + \gamma)} \,, 
\qquad q=e^{-\eta}  \,, 
\ee
as well as the $q$-analogue of the duplication formula \cite{GR}
\be
(1 + q)^{2 x -1}\ \Gamma_{q^{2}}(x)\ \Gamma_{q^{2}}(x + {1\over 2}) =
\Gamma_{q}(2 x)\ \Gamma_{q^{2}}({1\over 2}) \,, 
\ee
where $\Gamma_{q}(x)$ is the $q$-analogue of the 
Euler gamma function.
 
We find \footnote{Evidently, Eq. (\ref{openrel}) determines the product
$\alpha^{+}_{(l)} \alpha^{-}_{(l)}$. Further conditions are necessary 
to determine $\alpha^{-}_{(l)}$ and $\alpha^{+}_{(l)}$ separately. 
Following \cite{GMN}, we assume that the parts of $\alpha^{-}_{(l)}$ 
and $\alpha^{+}_{(l)}$ which do not originate from 
$\delta\sigma^{(1)}_{(l)}$ are equal.}
\be
\alpha^{-}_{(l)} &=& q^{- i 2 \tilde\lambda^{(1)} (1 + l)/\n}
S_{0}(\tilde\lambda^{(1)}) \non \\
&\times & {\Gamma_{q^{2\n}} \left({1\over \n}\left(\xi_{-} + l 
- {1\over 2} + i \tilde\lambda^{(1)} \right) \right) 
\Gamma_{q^{2\n}} \left({1\over \n}\left(\xi_{-} + \n - {1\over 2}
- i \tilde\lambda^{(1)} \right) \right)\over
\Gamma_{q^{2\n}} \left({1\over \n}\left(\xi_{-} + l - {1\over 2}
- i \tilde\lambda^{(1)} \right) \right) 
\Gamma_{q^{2\n}} \left({1\over \n}\left(\xi_{-} + \n - {1\over 2}
+ i \tilde\lambda^{(1)} \right) \right)} \,, \non \\
\alpha^{+}_{(l)} &=& q^{- i 2 \tilde\lambda^{(1)} (1 - l + 2 \n)/\n}
S_{0}(\tilde\lambda^{(1)}) \non \\
&\times  & {\Gamma_{q^{2\n}} \left({1\over \n}\left(\xi_{+} - \n + l 
- {1\over 2} - i \tilde\lambda^{(1)} \right) \right) 
\Gamma_{q^{2\n}} \left({1\over \n}\left(\xi_{+} - {1\over 2}
+ i \tilde\lambda^{(1)} \right) \right)\over
\Gamma_{q^{2\n}} \left({1\over \n}\left(\xi_{+} - \n + l - {1\over 2}
+ i \tilde\lambda^{(1)} \right) \right) 
\Gamma_{q^{2\n}} \left({1\over \n}\left(\xi_{+} - {1\over 2}
- i \tilde\lambda^{(1)} \right) \right)} \,,
\label{alphas}
\ee
where the prefactor $S_{0}(\tilde\lambda)$ is given by
\be
S_{0}(\tilde\lambda) = {\Gamma_{q^{4\n}} \left({1\over \n}\left({1\over 2}(\n -1) 
- i \tilde\lambda \right) \right) 
\Gamma_{q^{4\n}} \left({1\over \n}\left(\n + i \tilde\lambda \right) \right)\over
\Gamma_{q^{4\n}} \left({1\over \n}\left({1\over 2}(\n -1) 
+ i \tilde\lambda \right) \right) 
\Gamma_{q^{4\n}} \left({1\over \n}\left(\n - i \tilde\lambda \right) \right)}
\,. \label{prefactor}
\ee 
Furthermore, choosing the state $|\tilde\lambda^{(1)} \rangle$ in Eq.  
(\ref{quantizationopen}) to be the BA state constructed with 
any of the last $\n - l$ pseudovacua $\omega_{(l+1)} \,, \ldots \,, 
\omega_{(\n)}$ having one hole in sea 1, we obtain the relation
\be
{\beta^{+}_{(l)}\ \beta^{-}_{(l)}\over \alpha^{+}_{(l)}\ 
\alpha^{-}_{(l)}} = 
\exp \left\{ i 2\pi N \int_{0}^{\tilde\lambda^{(1)}}
\left( \sigma'^{(1)}_{(l)}(\lambda) - \sigma^{(1)}_{(l)}(\lambda)
\right) d\lambda \right\}  \,.
\ee 
We note that
\be
\sigma'^{(1)}_{(l)}(\lambda) - \sigma^{(1)}_{(l)}(\lambda) &=&
\delta\sigma'^{(1)}_{(l)}(\lambda) - 
\delta\sigma^{(1)}_{(l)}(\lambda) \non \\ 
&=& {1\over N} \left( a_{2\xi_{-}-1}(\lambda) 
- a_{2\xi_{+} - 2\n + 2l - 1}(\lambda) \right) \,, 
\ee 
where $\delta\sigma^{(1)}_{(l)}$ and $\delta\sigma'^{(1)}_{(l)}$ are 
given by Eqs.  (\ref{deltasigma}) and (\ref{deltasigmaprime}), 
respectively. We conclude
\be
{\alpha^{-}_{(l)}\over \beta^{-}_{(l)}} &=& 
-e_{2\xi_{-} - 1}(\tilde\lambda^{(1)}) \,, \non \\ 
{\beta^{+}_{(l)}\over \alpha^{+}_{(l)}} &=& 
-e_{2\xi_{+} - 2\n + 2l - 1}(\tilde\lambda^{(1)}) \,,
\label{betas}
\ee
where we have resolved the sign ambiguity by demanding that the $S$ 
matrix be proportional to the unit matrix for $\tilde\lambda^{(1)} = 0$.

We remark that there are relations among the boundary $S$ matrices. 
Indeed, the above results for  $S^{\mp}_{(l)\ [1]}$  are valid in the regime 
$\xi_{-} > {1\over 2}(\n -1)$, $\xi_{+} > \n - {1\over 2}$, to which 
we now refer as regime I. Corresponding results for regime II defined as
$\xi_{-} < -{1\over 2}(\n -1)$, $\xi_{+} < -\n + {1\over 2}$ can be 
obtained by a similar analysis, starting with the BAE (\ref{openBAE}) and 
replacing $\xi_{-} \rightarrow - \xi_{-} \,, \quad 
\xi_{+} \rightarrow - \xi_{+} + \n $. We observe that the $S$ matrices 
in these two regimes are related by duality:
\be
\alpha^{\mp\ (I)}_{(l)}\Big\vert_{\xi_{\mp} \rightarrow \xi'_{\mp} \,, 
\ l \rightarrow l'} &=& \beta^{\mp\ (II)}_{(l)} \non \\
\beta^{\mp\ (I)}_{(l)}\Big\vert_{\xi_{\mp} \rightarrow \xi'_{\mp} \,, 
\ l \rightarrow l'} &=& \alpha^{\mp\ (II)}_{(l)} \,,
\ee
where $\xi_{\pm}'$ and $l'$ are given by Eq. (\ref{primes}). 
Moreover, there is a relation between $S^{-}_{(l)\ [1]}$ and
$S^{+}_{(l)\ [1]}$,
\be
S^{-\ (I)}_{(l)\ [1]} \Big\vert_{\xi_{-} 
\rightarrow -\xi_{+} + \n - l} = 
S^{+\ (II)}_{(l)\ [1]} \,.
\ee 

Finally, we consider the case $[\n-1]$. The boundary $S$ matrices 
$S^{\mp}_{(l)\ [\n-1]}$ are diagonal $\n \times \n$ matrices of the 
form
\be
S^{\mp}_{(l)\ [\n-1]} = diag \Bigl(  
\underbrace{\bar \alpha^{\mp}_{(l)}\,, \ldots \,, 
\bar \alpha^{\mp}_{(l)}}_{l} \,, 
\underbrace{\bar \beta^{\mp}_{(l)}\,, \ldots \,, 
\bar \beta^{\mp}_{(l)}}_{\n-l} 
\Bigr) \,.
\label{form[n-1]}
\ee
For this case we must consider one hole in sea $\n - 1$. Noting that 
\be
\bar\beta^{+}_{(l)}\ \bar\beta^{-}_{(l)} \sim   
\exp \left\{ i 2\pi N \int_{0}^{\tilde\lambda^{(\n -1)}}
\left( \sigma_{(l)}^{(\n -1)}(\lambda) - 2 s^{(\n -1)}(\lambda) \right) 
d\lambda \right\} \,,
\ee 
we obtain
\be
\bar\beta^{-}_{(l)} &=& q^{- i 2 \tilde\lambda^{(\n-1)} (1 - l +\n)/\n}
S_{0}(\tilde\lambda^{(\n -1)}) \non \\
&\times  & {\Gamma_{q^{2\n}} \left({1\over \n}\left(\xi_{-} + l 
+ {1\over 2}(\n -1) - i \tilde\lambda^{(\n -1)} \right) \right)
\Gamma_{q^{2\n}} \left({1\over \n}\left(\xi_{-} + {1\over 2}(\n -1)
+ i \tilde\lambda^{(\n -1)} \right) \right) \over
\Gamma_{q^{2\n}} \left({1\over \n}\left(\xi_{-} + l + {1\over 2}(\n -1)
+ i \tilde\lambda^{(\n -1)} \right) \right)
\Gamma_{q^{2\n}} \left({1\over \n}\left(\xi_{-} + {1\over 2}(\n -1)
- i \tilde\lambda^{(\n -1)} \right) \right)} \,, \non \\
\bar\beta^{+}_{(l)} &=& q^{- i 2 \tilde\lambda^{(\n-1)} (1 + l +\n)/\n}
S_{0}(\tilde\lambda^{(\n -1)}) \non \\
&\times  & {\Gamma_{q^{2\n}} \left({1\over \n}\left(\xi_{+} + l 
- {1\over 2}(\n +1) + i \tilde\lambda^{(\n -1)} \right) \right)
\Gamma_{q^{2\n}} \left({1\over \n}\left(\xi_{+} - {1\over 2}(\n +1)
- i \tilde\lambda^{(\n -1)} \right) \right) \over
\Gamma_{q^{2\n}} \left({1\over \n}\left(\xi_{+} + l - {1\over 2}(\n +1)
- i \tilde\lambda^{(\n -1)} \right) \right)
\Gamma_{q^{2\n}} \left({1\over \n}\left(\xi_{+} - {1\over 2}(\n +1)
+ i \tilde\lambda^{(\n -1)} \right) \right)} \,, \non \\
& & 
\label{baralphas}
\ee
where $S_{0}(\tilde\lambda) $ is given in Eq. (\ref{prefactor}). 
Moreover,
\be
{\bar\beta^{-}_{(l)}\over \bar\alpha^{-}_{(l)}} &=& 
-e_{2\xi_{-} + 2l - \n - 1}(\tilde\lambda^{(\n -1)}) \,, \non \\ 
{\bar\alpha^{+}_{(l)}\over \bar\beta^{+}_{(l)}} &=& 
-e_{2\xi_{+} - \n - 1}(\tilde\lambda^{(\n -1)}) \,.
\ee

For the case $\n=2$, we recover the results of Refs.  \cite{JKKKM} and 
\cite{DMN2}.  In the isotropic limit $\eta \rightarrow 0$, we see that 
$q \rightarrow 1$ and $\Gamma_{q}(x) \rightarrow \Gamma(x)$; hence, we 
recover the recent results \cite{DN}.

\section{Discussion}

The transfer matrix $t_{(l)}(\lambda \,, \xi_{-} \,, \xi_{+})$ has the 
symmetry $U_{q}\left( SU(l) \right)$ $\times$ $U_{q}\left( SU(\n-l) 
\right)$ $\times$ $U(1)$ as well as a duality symmetry $l 
\leftrightarrow \n - l$, as we have shown in Eqs.  (\ref{first}), 
(\ref{second}) and (\ref{duality}), respectively.  It should be 
possible to show, generalizing \cite{mn/mpla}, \cite{DVGR2}, 
\cite{highestweight}, that the corresponding Bethe Ansatz states $| \ 
\rangle$ are highest weights of $U_{q}\left( SU(l) \right) \times 
U_{q}\left( SU(\n-l) \right)$.  That is,
\be
J^{+(k)} |\ \rangle = 0 \,, \qquad k \ne l \,.
\ee
This symmetry was used here for computing boundary $S$ matrices. We 
expect that this symmetry will also be valuable for computing other 
quantities of physical interest (e.g, form factors, correlation 
functions, etc.) and for elucidating the phase structure of the 
corresponding vertex model. Although we have considered here the 
noncritical regime, we expect that this approach can be extended to 
the critical regime $|q|=1$.

The ``mixed'' boundary condition case $l_{+} \ne l_{-}$ merits further 
investigation.  One expects that the boundary $S$ matrix for one end 
of the chain should be independent of the boundary conditions at the 
other end.  However, for mixed boundary conditions, the unbroken 
symmetry group of the transfer matrix is smaller, and therefore, the 
arguments presented here require further refinement.

We have seen that by turning on diagonal boundary interactions, the 
symmetry algebra of the transfer matrix is broken to a nontrivial 
subalgebra.  We expect that this is a general phenomenon.  More 
explicitly, consider the $R$ matrix associated with the affine Lie 
algebra $\hat g$ \cite{bazhanov}, \cite{jimbo}.  It is known 
\cite{mn/mpla} that the corresponding open spin chain transfer matrix 
with $K = 1$ (i.e., without additional boundary interactions) has the 
symmetry $U_{q}\left( g_{0} \right)$, where $g_{0}$ is the maximal 
finite-dimensional subalgebra of $\hat g$ .  We expect that the 
transfer matrix with diagonal $K \ne 1$ will have as its symmetry a 
nontrivial subalgebra of $U_{q}\left( g_{0} \right)$.

We emphasize that we have considered here only {\it diagonal} boundary 
interactions.  It would be interesting to look for symmetries of the 
transfer matrix with nondiagonal $K$ matrices.  Unfortunately, even 
for the simplest case $A_{1}^{(1)}$, the Bethe Ansatz solution is not 
yet known, which precludes computation of the boundary $S$ matrix in 
the manner described here.

We also emphasize that here we have restricted our attention to exact 
symmetries of the {\it finite} chain. For the (semi) infinite chain, 
one expects additional affine symmetries \cite{JKKKM},\cite{JM}.
In field theory, such symmetries are generated by fractional-spin 
integrals of motion \cite{luscher},\cite{sasha},\cite{BL}, which can
exist even for nondiagonal boundary interactions \cite{mn/bsg}. 

\section{Acknowledgments}

We are grateful to O.  Alvarez and V.  Brazhnikov for valuable 
discussions, and to L.  Mezincescu for prior collaborations.  This 
work was supported in part by the National Science Foundation under 
Grants PHY-9507829 and PHY-9870101.

\appendix
\section{Proof of Eqs. (\ref{finalFirst}) and (\ref{finalSecond})}
	
We prove here the relations (\ref{finalFirst}), (\ref{finalSecond}) 
which enter our proof of quantum-algebra symmetry of the transfer 
matrix. 

We begin by noting the identities
\be
\Pi_{(l)\ 1}^{\pm}\ R_{12}^{\pm}\ \Pi_{(l)\ 1}^{\pm} = 
R_{12}^{\pm}\ \Pi_{(l)\ 1}^{\pm} \,,
\ee
which imply
\be
\Pi_{(l)\ 1}^{\pm}\ R_{21}^{\pm}\ \Pi_{(l)\ 1}^{\pm} = 
\Pi_{(l)\ 1}^{\pm}\ R_{21}^{\pm} \,,
\ee
and therefore
\be
\Pi_{(l)\ 0}^{\pm}\ T^{\pm}_{0}\ \Pi_{(l)\ 0}^{\pm} &=& 
T^{\pm}_{0}\ \Pi_{(l)\ 0}^{\pm} \,, \non \\ 
\Pi_{(l)\ 0}^{\pm}\ \hat T^{\pm}_{0}\ \Pi_{(l)\ 0}^{\pm} &=& 
\Pi_{(l)\ 0}^{\pm}\ \hat T^{\pm}_{0}\  \,.
\ee
It is now easy to show that 
\be
\Pi_{(l)\ 0}^{\pm}\ {\cal T}_{(l)\ 0}^{\pm}\ \Pi_{(l)\ 0}^{\pm} =
{\cal T}_{(l)\ 0}^{\pm} \,.
\label{OrthogSubspaces}
\ee 
Indeed, recalling Eq. (\ref{calTpm}),
\be
\Pi_{(l)\ 0}^{\pm}\ {\cal T}_{(l)\ 0}^{\pm}\ \Pi_{(l)\ 0}^{\pm} =
\Pi_{(l)\ 0}^{\pm}\ T_{0}^{\pm}\ \Pi_{(l)\ 0}^{\pm}\ \hat T_{0}^{\pm}\ 
\Pi_{(l)\ 0}^{\pm} 
= T_{0}^{\pm}\ \Pi_{(l)\ 0}^{\pm}\ \hat T_{0}^{\pm} ={\cal T}_{(l)\ 0}^{\pm} 
\,.
\ee
In a similar manner one can prove that
\be
\Pi_{(l)\ 1}^{\pm} \left( R_{12}^{\pm}\ {\cal T}_{(l)\ 1}^{\pm}\
R_{21}^{\pm}\ {\cal T}_{(l)\ 2}(\lambda \,, \xi ) \right) \Pi_{(l)\ 1}^{\pm} =
R_{12}^{\pm}\ {\cal T}_{(l)\ 1}^{\pm}\
R_{21}^{\pm}\ {\cal T}_{(l)\ 2}(\lambda \,, \xi ) 
\,.
\ee 
This relation, together with the fact
\be
\Pi_{(l)}^{\pm}\ M\ \Pi_{(l)}^{\pm} = \Pi_{(l)}^{\pm}\ M = M\ 
\Pi_{(l)}^{\pm} \,,
\ee
imply the first desired result (\ref{finalFirst}).

We next note the identities
\be
\Pi_{(l)\ 1}^{\pm}\ R_{12}^{\mp}\ \Pi_{(l)\ 1}^{\pm} = 
\Pi_{(l)\ 1}^{\pm}\ R_{12}^{\mp}  \,,
\ee
which imply
\be
\Pi_{(l)\ 1}^{\pm}\ R_{21}^{\mp}\ \Pi_{(l)\ 1}^{\pm} = 
R_{21}^{\mp}\ \Pi_{(l)\ 1}^{\pm}\,.
\ee

Let us denote by $R'^{\mp}$ the ``projected'' $R$ matrix, i.e., 
\be
R'^{\mp}_{12} = \Pi_{(l)\ 1}^{\pm}\ R_{12}^{\mp}\ \Pi_{(l)\ 1}^{\pm} \,.
\ee
It is easy to see that this projected $R$ matrix commutes with 
$K_{(l)\ 2}(\lambda \,, \xi)$,
\be
\left[ R'^{\mp}_{12} \,, K_{(l)\ 2}(\lambda \,, \xi) \right] = 0 \,.
\label{littlelemma}
\ee
Indeed, we see from Eq. (\ref{Kmatrix}) that
\be
K_{(l)}(\lambda \,, \xi) = a \Pi_{(l)}^{+} + b \Pi_{(l)}^{-} \,.
\ee
Since $\left[ R_{12}^{-} \,, \Pi_{(l)\ 1}^{\pm} \Pi_{(l)\ 2}^{\pm} 
\right] = 0$, then 
$\left[ R'^{-}_{12} \,, \Pi_{(l)\ 2}^{\pm} \right] = 0$, and the 
result (\ref{littlelemma}) for $R'^{-}_{12}$ immediately follows. The 
proof for $R'^{+}_{12}$ is obtained in similar fashion.

All the pieces are now in place for proving the relation
\be
\Pi_{(l)\ 1}^{\pm} \left( R_{12}^{\mp} P_{1}^{\pm t_{1}} M_{1}^{-1} 
R_{21}^{\mp} K_{(l)\ 2}(\lambda \,, \xi) \right) \Pi_{(l)\ 1}^{\pm} 
= \Pi_{(l)\ 1}^{\pm} \left( K_{(l)\ 2}(\lambda \,, \xi) 
R^{\mp}_{12} P_{1}^{\pm \ t_{1}} M_{1}^{-1} R_{21}^{\mp} \right) 
\Pi_{(l)\ 1}^{\pm} \,, \non \\
\ee 
which is the second desired result (\ref{finalSecond}). Indeed,
\be
LHS = R'^{\mp}_{12} P_{1}^{\pm t_{1}} M_{1}^{-1} R'^{\mp}_{21}
K_{(l)\ 2}(\lambda \,, \xi) 
= K_{(l)\ 2}(\lambda \,, \xi) R'^{\mp}_{12} P_{1}^{\pm t_{1}} 
M_{1}^{-1} R'^{\mp}_{21} = RHS \,.
\ee

\section{Bethe Ansatz Equations for general pseudovacua}

A priori, one expects that the Bethe Ansatz equations (BAE) depend on 
the choice of pseudovacuum.
We show here that the BAE for the transfer matrix $t_{(l)}(\lambda \,, 
\xi_{-} \,, \xi_{+})$ corresponding to the pseudovacua $\omega_{(1)} 
\,, \ldots \,, \omega_{(l)}$ are all the same; and that the BAE 
corresponding to the pseudovacua $\omega_{(l+1)} \,, \ldots \,, 
\omega_{(\n)}$ are all the same.

We first observe that the operators $\left( J^{\pm (k)}\right)^{N}$, 
where $N$ is the number of sites, transform one pseudovacuum into 
another:
\be
\left( J^{- (k)}\right)^{N}\ \omega_{(j)} &=& \omega_{(j+1)}\ \delta_{k 
\,, j} \,, \non \\
\left( J^{+ (k)}\right)^{N}\ \omega_{(j)} &=& \omega_{(j-1)}\ \delta_{k 
\,, j-1} \,.
\ee
The $U_{q}\left( SU(l) 
\right)$ $\times$ $U_{q}\left( SU(\n-l) \right)$ symmetry of the 
transfer matrix (\ref{second}) implies the (weaker) result
\be
\left[ t_{(l)}(\lambda \,, \xi_{-} \,, \xi_{+}) 
\,, \left( J^{\pm (k)}\right)^{N} 
\right] =
0 \,, \qquad k \ne l \,.
\ee
Let $\Lambda^{(0)}_{(l)}(\lambda \,, \xi_{-} \,, \xi_{+})$ denote the 
eigenvalue of the transfer matrix corresponding to the first
pseudovacuum $\omega_{(1)}$,
\be
t_{(l)}(\lambda \,, \xi_{-} \,, \xi_{+})\ \omega_{(1)}
= \Lambda^{(0)}_{(l)}(\lambda \,, \xi_{-} \,, \xi_{+})\ \omega_{(1)}
\,.
\ee
Acting repeatedly on both sides of this equation with the lowering 
operator $\left( J^{- (k)}\right)^{N}$, we conclude that the transfer 
matrix has the same eigenvalue with each of the first $l$ pseudovacua,
\be
t_{(l)}(\lambda \,, \xi_{-} \,, \xi_{+})\ \omega_{(k)}
= \Lambda^{(0)}_{(l)}(\lambda \,, \xi_{-} \,, \xi_{+})\ \omega_{(k)}
\,, \qquad k=1 \,, \ldots \,, l \,.
\ee
Similarly, let $\Lambda'^{(0)}_{(l)}(\lambda \,, \xi_{-} \,, \xi_{+})$ 
be the eigenvalue of the transfer matrix corresponding to the last
pseudovacuum $\omega_{(\n)}$,
\be
t_{(l)}(\lambda \,, \xi_{-} \,, \xi_{+})\ \omega_{(\n)}
= \Lambda'^{(0)}_{(l)}(\lambda \,, \xi_{-} \,, \xi_{+})\ \omega_{(\n)}
\,.
\ee
Acting on both sides with the raising operator 
$\left( J^{+ (k)}\right)^{N}$, we see that the transfer matrix 
has the same eigenvalue with each of the last $\n-l$ pseudovacua,
\be
t_{(l)}(\lambda \,, \xi_{-} \,, \xi_{+})\ \omega_{(k)}
= \Lambda'^{(0)}_{(l)}(\lambda \,, \xi_{-} \,, \xi_{+})\ \omega_{(k)}
\,, \qquad k=l+1 \,, \ldots \,, \n \,.
\ee

We have shown so far that the {\it vacuum} eigenvalues of 
$t_{(l)}(\lambda \,, \xi_{-} \,, \xi_{+})$ with  
$\omega_{(1)} \,, \ldots \,, \omega_{(l)}$ are the same, and those 
with $\omega_{(l+1)} \,, \ldots \,, \omega_{(\n)}$ are the same.
The eigenvalues of $t_{(l)}(\lambda \,, \xi_{-} \,, \xi_{+})$  
corresponding to {\it general} BA states are ``dressed'' pseudovacuum 
eigenvalues, with dressing functions that can be deduced from 
properties of the transfer matrix such as analyticity, fusion, etc., 
independently of the choice of pseudovacuum. (See, e.g., 
\cite{reshetikhin}, \cite{mn/npb}.) It follows that the 
expressions for the eigenvalues of the transfer matrix corresponding 
to general BA states constructed with pseudovacua 
$\omega_{(1)} \,, \ldots \,, \omega_{(l)}$ 
are all the same, and analogously for the pseudovacua
$\omega_{(l+1)} \,, \ldots \,, \omega_{(\n)}$. Since the BAE are the 
conditions that the eigenvalues have vanishing residues, we conclude
that the BAE corresponding to the pseudovacua $\omega_{(1)} 
\,, \ldots \,, \omega_{(l)}$ are all the same; and that the BAE 
corresponding to the pseudovacua $\omega_{(l+1)} \,, \ldots \,, 
\omega_{(\n)}$ are all the same.

We remark that the above proof could be streamlined (in 
particular, avoiding the argument of dressing the pseudovacuum 
eigenvalues) if one could construct invertible transformations 
${\cal U}_{(j)}$ which keep the transfer matrix invariant
and change the pseudovacuum,
\be
{\cal U}_{(j)}\ t_{(l)}(\lambda \,, \xi_{-} \,, \xi_{+})\ 
{\cal U}^{-1}_{(j)} &=& t_{(l)}(\lambda \,, \xi_{-} \,, \xi_{+}) \,, 
\non \\
{\cal U}_{(j)}\ \omega_{(j)} &=& \omega_{(j + 1)} \,,
\ee
for $j \ne l$.
For the isotropic case, such transformations are finite elements of 
the groups $SU(l)$ and $SU(\n-l)$, which can easily be constructed 
\cite{DN}.  Unfortunately, for the anisotropic case, we have not yet 
succeeded in constructing such transformations for general values of 
the number of sites, $N$ .  Such transformations are presumably 
related to the so-called $q$-Weyl group \cite{KR}, and may have other 
interesting applications.

\section{Symmetry transformations which change the anisotropy 
parameter}

The symmetries of the transfer matrix discussed in Section 2 leave the 
anisotropy parameter $\eta$ unchanged. We briefly discuss here related 
transformations, which however transform $\eta \rightarrow -\eta$. Let 
us now make the dependence on $\eta$ explicit, and  
denote the $R$ matrix (\ref{Rmatrix}) by $R_{12}(\lambda \,, 
\eta)$, and the $K$ matrix (\ref{Kmatrix}) by $K_{(l)}(\lambda \,, 
\xi \,, \eta)$. Evidently,
\be
R_{21}(\lambda \,, \eta) = R_{12}(\lambda \,, -\eta) \,.
\ee
Moreover, let $U$ now be the antidiagonal $\n \times \n$ matrix
\be
U = \left( \begin{array}{cccc}
	& &  1      \\
	& \lddots    \\
	1
\end{array} \right) \,.
\ee 
One can show that that
\be
U_{1}\ U_{2}\ R_{12}(\lambda \,, \eta)\ U_{1}^{-1}\ U_{2}^{-1} &=&
 R_{12}(\lambda \,, -\eta) \,, \non \\
U\ K_{(l)}(\lambda \,, \xi \,, \eta)\ U^{-1} &=&
K_{(l')}(\lambda \,, \xi' \,, -\eta) \,,
\ee
where $l'$ and $\xi_{\pm}'$ are given by Eq. (\ref{primes}).
The transfer matrix transforms under the corresponding quantum-space 
transformation ${\cal U}$ as follows:
\be
{\cal U}\ t_{(l)}(\lambda \,, \xi_{-} \,, \xi_{+}\,, \eta )\ 
{\cal U}^{-1} \propto 
t_{(l')}(\lambda \,, \xi_{-}' \,, \xi_{+}'\,, -\eta ) \,.
\ee
Like the duality transformation (\ref{duality}), this transformation 
maps $l \rightarrow l'$ and $\xi_{\pm} \rightarrow \xi'_{\pm}$ , but 
it also maps $\eta \rightarrow - \eta$.

Furthermore, define the two $\n \times \n$ matrices
\be
U_{(l)}^{+} = \left(
\begin{array}{cccccc}
	& & 1      \\
	& \lddots   \\
	1          \\
	& & & 1    \\
	& & & & \ddots \\
	& & & & & 	1
\end{array} \right) \,, \qquad 
U_{(l)}^{-} = \left(
\begin{array}{cccccc}
	1              \\
	& \ddots       \\
	& & 1          \\
	& & & & &  1   \\
	& & & & \lddots \\
	& & & 1
\end{array} \right) \,.
\ee 
with an antidiagonal upper $l \times l$ block, and 
an antidiagonal lower $(\n - l) \times (\n - l)$ block,
respectively. The 
corresponding quantum-space transformations ${\cal U}^{\pm}_{(l)}$ 
have the following action on the transfer matrix
\be
{\cal U}^{\pm}_{(l)}\ t_{(l)}(\lambda \,, \xi_{-} \,, \xi_{+}\,, \eta )\ 
{\cal U}^{\pm\ -1}_{(l)} \propto 
t_{(l)}(\lambda \,, \xi_{-} \,, \xi_{+}\,, -\eta ) \,.
\ee
However, for $\n > 3$, there are not enough of these transformations 
to determine all the elements of the boundary $S$ matrices.

\end{document}